\documentclass[aps,prd,showpacs]{revtex4}
\usepackage{amsmath}
\usepackage[english]{babel}
\usepackage[latin1]{inputenc}
\usepackage{amssymb}
\usepackage{graphicx}
\usepackage{natbib}
\usepackage{epsfig}
\usepackage{bm}
\usepackage{dcolumn}

\usepackage[dvipsnames]{xcolor}

\begin{document}

\vspace*{2cm}

\title{On the
pentaquark candidates $P_c^+(4380)$ and $P_c^+(4450)$ within the
soliton picture of baryons}

\author{N.N. Scoccola$^{a,b,c}$, D.O. Riska$^{d}$ and Mannque Rho$^{e}$ }

\affiliation{
$^a$ Department of Theoretical Physics, Comisi\'on Nacional de
Energ\'ia
At\'omica, Av.Libertador 8250, 1429 Buenos Aires, Argentina\\
$^b$ CONICET, Rivadavia 1917, 1033 Buenos Aires, Argentina\\
$^c$ Universidad Favaloro, Sol\'is 453, 1078 Buenos Aires,
Argentina\\
$^d$ Finnish Society of Science and Letters, Fabianinkatu 4 B 16,
00130 Helsinki, Finland
\\
$^e$Institut de Physique Th\'{e}orique, CEA Saclay,
91191 Gif-sur-Yvette, France
}


\begin{abstract}
Using the bound state version of the topological soliton model for the
baryons we show that the existence of a bound (or quasi-bound) $\bar D$-soliton state
leads to the possibility of having hidden charm pentaquarks with quantum numbers and
masses, which are compatible with those of the candidates recently reported by the LHCb
experiment. The implications of heavy quark symmetry are elaborated.
\end{abstract}

\maketitle

\vspace*{1cm}

\section{Introduction}

The LHCb collaboration at CERN recently reported the discovery of
two states, which in the quark model correspond to ``pentaquarks" with
hidden charm Ref.\cite{Aaij:2015tga}. The first one has the mass (width) $4380 ± 8 ± 29$
($205\pm 18\pm 86$ ) MeV and the
second the mass (width) $4449.8 ± 1.7 ± 2.5$ ( $39\pm 5\pm 19$ ) MeV.  The preferred quantum
number assignments are $J^\pi = 3/2^-, 5/2^+$, respectively,
although acceptable solutions are also found for additional cases
with opposite parity. Note that their decay channel is $J/\psi \
p^+$ which implies $I=1/2$.
After this announcement several articles that consider possible theoretical
interpretations of the observed states have appeared. Some of these
\cite{Chen:2015loa,Chen:2015moa,Roca:2015dva,He:2015cea}
suggest, with some variations, that the observed states may be interpreted as
anticharmed meson ($\bar D$ or $\bar D^*$)-hyperon ($\Sigma$ or $\Sigma^*$) molecular states, while
others base their description on diquark models \cite{Maiani:2015vwa,Lebed:2015tna,Anisovich:2015cia}.
The possibility that at least one of the observed peaks might be only a kinematical
effect was discussed in Refs.\cite{Guo:2015umn,Liu:2015fea,Mikhasenko:2015vca}.
Further suggestions have been put forward in Refs.\cite{Mironov:2015ica,Meissner:2015mza}.
Moreover, the
production and formation of the hidden-charm pentaquarks in $\gamma$ - nucleon collisions
have been discussed in Refs.\cite{Kubarovsky:2015aaa,Wang:2015jsa}.

The purpose of the present note is to show that it is also
possible to account for the quantum numbers and masses of these
observed pentaquark candidates within the topological soliton
picture of baryons. The possible existence of stable pentaquarks
with negative charm in that framework was already considered many
years ago \cite{Riska:1992qd}. In this approach the heavy flavor
hyperons were described as bound states of heavy flavour mesons
and a topological soliton
\cite{Rho:1992yy,Riska:1992qd,Riska:1991qd} in the extension to
heavy flavor of the bound state
approximation \cite{Callan:1985hy,Callan:1987xt,Scoccola:1988wa} to
the Skyrme model, supplemented with suitable symmetry breaking
terms. This ``Naive Skyrme Model" (NSM) formulation does however
only approximately \cite{Bjornberg:1991hc} incorporate heavy quark
symmetry (HQS) according to which in the heavy quark limit the
heavy pseudoscalar and vector fields become degenerate and, therefore
should be treated on an equal footing \cite{Manohar:2000dt}. An
improved way to proceed is therefore to apply the bound state
approach to the heavy meson effective lagrangian
\cite{Wis92,BD92,Yan92,Goity:1992tp}, which simultaneously
incorporates chiral symmetry and heavy quark symmetry (for details
see Ref.\cite{Scoccola:2009au} and refs. therein). The possible
existence of  a $C=-1$ meson bound state in the context of a model
consistent with HQS was first discussed in
Refs.\cite{Oh:1994np,Oh:1994ux}. There, however, only pseudoscalar
mesons as described by the Skyrme model were considered in the
light sector. In what follows we will refer to this as SMHQS
formulation. It was later pointed out \cite{Schechter:1995vr}
that the inclusion of light
vector mesons in the corresponding effective lagrangian tends to
push this state into the continuum.
On other hand, the calculation in Ref.\cite{Oh:1997tp}, which
incorporates the center of mass corrections in a more consistent
way, still leaves the possibility of a loosely bound state.
Additional arguments for the existence of $C=-1$ meson-soliton
bound state have been given in Ref.\cite{Harada:2012dm}.

In the NSM approach it is the Wess-Zumino term in the lagrangian,
which is responsible for the difference in the interaction of the
soliton and the mesons with opposite massive flavor quantum
number. This term is repulsive in the case of mesons with massive
antiflavor quantum number. In the case of $S=+1$ kaons, this
repulsion, in combination with the repulsive effect of the meson
kinetic energy term, pushes them into the continuum
\cite{Scoccola:1990pt,Itzhaki:2003nr}. (It should here be
mentioned that the indications for the existence of the
conjectured strange pentaquark ``$\theta(1540)$"
\cite{Diakonov:1997mm} have hitherto not been experimentally
confirmed\cite{Hicks:2012zz}). As the repulsive effect of the
kinetic energy term weakens with increasing meson mass and the
strength of the Wess-Zumino term is smaller for heavy flavors, the
existence of anticharm (and {\it a fortiori} antibottom)
meson-soliton bound states becomes possible. Below we show that
the existence of a bound (or quasi-bound state) $\bar D$-soliton
state naturally leads to the possibility of having some hidden
charm pentaquarks with quantum numbers and masses which are
compatible with those of the candidates proposed in
Ref.\cite{Aaij:2015tga}

\section{Bound state description of hidden heavy flavoured pentaquarks}

Since the pentaquark candidates reported in
Ref.\cite{Aaij:2015tga} have no net charm quantum number we
propose to describe them as bound states of  a soliton and two
pseudoscalar mesons (one charm and the other anticharm) in the
present picture. We recall that in the bound state approximation,
while a bound $C=+1$ meson behaves as a quark
\cite{Rho:1992yy,Riska:1991qd,Callan:1985hy,Callan:1987xt,Scoccola:1988wa},
a $\bar D$-meson corresponds to a antiheavy-light quark pair
\cite{Riska:1992qd,Oh:1994np,Oh:1994ux}. To determine the possible
quantum numbers and estimate the values of the associated masses
we need to know which bound states of charm and anticharm mesons
are possible. In fact, both the NSM and the SMHQS models lead to a
number of meson bound states. Those bound states can be labelled
by $k^\pi$, where $\vec k = \vec i + \vec \ell$ and $\pi =
(-1)^{\ell + 1}$. Here, $\ell$ and $i=1/2$ are the angular
momentum and isospin of the bound pseudoscalar meson. In Table I
the corresponding quantum numbers, binding energies $b$ and
hyperfine splittings constants $c$ are given . The latter are
needed for calculation of the nonadiabatic corrections to be
discussed below. These results were obtained using the standard
Skyrme model parameters $f_\pi=64.5$ MeV, $e=5.45$, which lead to
the empirical masses for the nucleon and the $\Delta$ resonance.
The values of the binding energies $b$ and hyperfine splittings
$c$ associated with the $C=\pm 1$ mesons in the NSM scheme have
been calculated in Refs.\cite{Riska:1991qd,Riska:1992qd}. Those of
the SMHQS scheme have been extracted from Ref.\cite{Oh:1995ey}
(Set 5) except for the binding of the $C=-1$ state, which is taken
from Ref.\cite{Oh:1997tp}. Note that  most of the associated
hyperfine splittings have not been given in those works, however.
The results corresponding to the NSM have been obtained using the
decay constant ratio $f_D/f_\pi = 1.8$. This value has been
updated in recent years. The current estimate is $f_D/f_\pi =
1.57$ \cite{Agashe:2014kda}. While the use of this value hardly
affects the predictions for the hyperfine splitting constants, one
obtains an enhancement of about $100$ MeV for all the $C=+1$ binding
energies. Since the results for the binding energies obtained with
$f_D/f_\pi = 1.8$ are closer to those of the SMHQS formulation we
will assume that this overbinding is a consequence of
the simplicity inherent to the NSM formulation.

\vspace*{1cm}

\begin{table}[h]
\begin{center}
\begin{tabular}{ccc ccc c cc}
\hline
&&&\multicolumn{3}{c}{NSM}&&\multicolumn{2}{c}{SMHQS}   \\ \cline{5-6}\cline{8-9}
                         & $l$&$k^\pi$ &&  $b(MeV)$&$c$  &&   $b(MeV)$& $c$   \\
                         \hline
     $ C = +1 $          & 1 & $1/2^+$ && 568 & 0.20 && 518 & 0.15\\
                         & 0 & $1/2^-$ && 355 & 0.52 && 239 & 0.30 \\
                         & 2 & $3/2^-$ && 243 & 0.15 && 212 &  \\
                         & 1 & $3/2^+$ && 140 & 0.28 &&  49 &  \\
                         & 1 & $1/2^+$ && 118 & 0.03 &&  65 &  \\
\hline
     $ C = -1 $          & 1 & $1/2^+$ &&  38 & 0.16 && 54  &   \\[2.mm]
\hline
\end{tabular}
\end{center}
\caption{Quantum numbers of the meson bound states and associated binding energies
and hyperfine splitting constants.}
\end{table}

\vspace*{1cm}

The single meson spectra that follow from Table I are illustrated
in Fig.1.
It can be seen that the meson spectrum obtained in the NSM formulation
is qualitatively similar to that obtained using the SMHQS formulation.
Actually, this fact was already noticed long ago \cite{Bjornberg:1991hc}.

\begin{center}
\begin{figure}[hbt]
\includegraphics[width=0.7\textwidth]{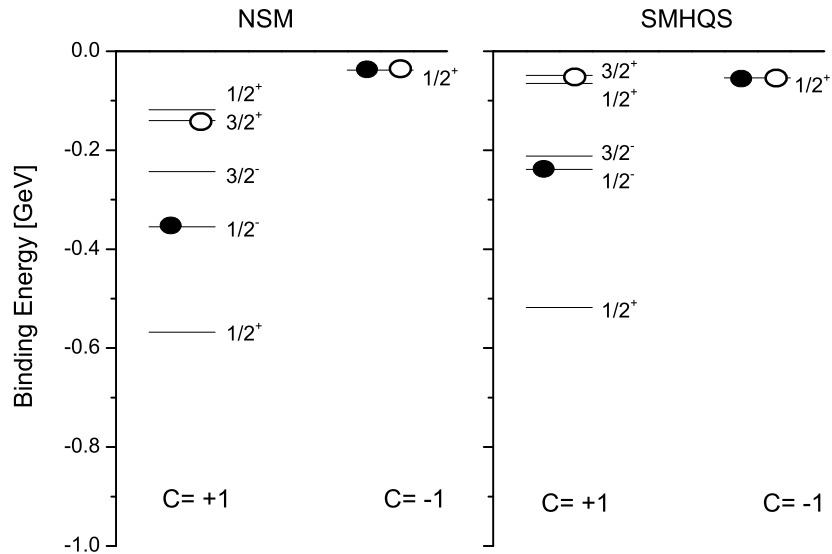}
\caption{Binding energies and quantum numbers corresponding to
possible soliton-(anti)charmed mesons. The left panel corresponds to
the spectrum obtained using the NSM formulation while the
right panel to the SMHQS one. The lowest $J^\pi = 3/2^-$
configuration is indicated by the two black circles and the $J^\pi = 5/2^+$ by
the open ones.}
\label{fig1}
\end{figure}
\end{center}
\vspace*{5mm}
Clearly, the $C=-1$ meson has to be bound in the
$k_-^{\pi_-}=1/2^{+}$ state. Here, the subindex stands for the
charm quantum number of the meson. For the $C=+1$ one, we have,
however, several possibilities. The preferred
quantum number assignments of the pentaquark candidates are $J^\pi
= 3/2^-, 5/2^+$, respectively. Since, obviously, $\pi = \pi_- \
\pi_+$ it follows that $\pi_+ = - (+)$ for the 3/2 (5/2) pentaquarks.
To determine $k_+$ we note that, according to the usual rules, the
total grand spin $\vec K= \vec k_+ + \vec k_-$ satisfies $\vec K =
\vec J + \vec I$ where $J$ and $I$ are the total spin and isopin
of the bound system. Noting that the observed states are isospin
doublets,
it follows that $1/2 \leq k_+ \leq 5/2$ for the
state with $J=3/2$ and $3/2 \leq k_+ \leq 7/2$ for that with $J=5/2$.
From Fig.1 it is clear that the lowest lying meson configurations
$(k_+^{\pi_+}, k_-^{\pi_-})$ that satisfy all the requirements are
$(1/2^-,1/2^+)$ for the $J^\pi=3/2^-$ pentaquark candidate and
$(3/2^+,1/2^+)$ for the $J^\pi=5/2^+$ one. In the figure the first
configuration is indicated by the two black circles and the second by
the open ones. In principle many other states can be obtained
by populating the alternative $C=+1$ bound states. The full list
is given in Table II together with the corresponding masses
up to non-adiabatic corrections. These have been obtained using
\begin{eqnarray}
M = M_{sol} + 2 m_D - b_+ - b_- ,
\end{eqnarray}
where $M_{sol}=866$ MeV is the soliton mass, $m_D$ the pseudoscalar charm
meson mass that we take to be $m_D = 1867$ MeV and $b_\pm$ are
the meson binding energies given in Table I.
\begin{table}[h]
\begin{center}
\begin{tabular}{ccc cccc}
  \hline
\hspace*{.2cm} I \hspace*{.2cm} &
\hspace*{.2cm} $k_+^{\pi_+}$ \hspace*{.2cm}&
\hspace*{.2cm} $k_-^{\pi_-}$ \hspace*{.2cm} &
\hspace*{.2cm} $K$ \hspace*{.2cm}    &
\hspace*{.2cm} $J^\pi$ \hspace*{.2cm} &
NSM & SMHQS
\\ \hline
1/2      &  $1/2^+$   &  $1/2^+$      &   0     & $1/2^+$ & 4.00 &  4.03  \\
         &            &               &   1     & $1/2^+$ &      &        \\
         &            &               &   1     & $3/2^+$ &      &        \\
\hline
1/2      &  $1/2^-$   &  $1/2^+$      &   0     & $1/2^-$ & 4.21 &  4.31  \\
         &            &               &   1     & $1/2^-$ &      &        \\
         &            &               &   1     & $3/2^-$ &      &        \\
\hline
1/2      &  $3/2^-$   &  $1/2^+$      &   1     & $1/2^-$ & 4.32 &  4.33  \\
         &            &               &   1     & $3/2^-$ &      &        \\
         &            &               &   2     & $3/2^-$ &      &        \\
         &            &               &   2     & $5/2^-$ &      &        \\
\hline
1/2      &  $3/2^+$   &  $1/2^+$      &   1     & $1/2^+$ & 4.42 &  4.50  \\
         &            &               &   1     & $3/2^+$ &      &        \\
         &            &               &   2     & $3/2^+$ &      &        \\
         &            &               &   2     & $5/2^+$ &      &        \\
\hline
1/2      &  $1/2^+$   &  $1/2^+$      &   0     & $1/2^+$ & 4.44 &  4.48  \\
         &            &               &   1     & $3/2^+$ &      &       \\
         &            &               &   1     & $3/2^+$ &      &       \\
\hline
\end{tabular}
\end{center}
\caption{\label{pnjl} Quantum numbers and ${\cal O}(N_c^0)$
masses (in GeV).}
\end{table}

It is seen that three low lying $3/2^-$ with masses (4.21-4.32,
4.31-4.33) GeV for (NSM,SMHQS), respectively, are predicted in
both schemes. The rather large width found in the experiment might
be explained by these 3 close lying states. One the other hand
only one single $5/2^+$ state at (4.42, 4.50) GeV is predicted.

So far we have not included the non-adiabatic contributions to the masses. To first order perturbation theory
the rotational corrections can be obtained by considering
\begin{eqnarray}
 {1 \over 2 \Omega} <(I (k_+ k_-)^{K})^J | (\vec R - \vec \Theta)^2| (I (k_+ k_-)^K)^J >,
\label{rotcorr}
\end{eqnarray}
where $\Omega$ is the moment of inertia of the soliton and
$\Theta$ the total isospin of the meson bound system. Moreover, $\vec R$ has the role
of spin of the light quark system (basically the spin of the rotating soliton, which coincides
with its isospin). For a system of a $C=+1$ and a $C=-1$ bound mesons one has
$\vec \Theta = \vec \theta_+ + \vec \theta_-$.
The calculation of the matrix element in Eq.(\ref{rotcorr}) requires some assumptions and/or
approximations. The usual procedure in the bound state approach (BSA) to the Skyrme model
\cite{Rho:1992yy,Riska:1991qd,Callan:1985hy,Callan:1987xt,Scoccola:1988wa}
is to employ the approximation $<\theta_i^2> = c_i^2 k_i (k_i+1)$. We denote this as BSA option.
Here, $c_i$ is the corresponding hyperfine splitting constant.
As noted in Ref.\cite{Oh:1994yv}
this approximation does not hold in the HQS limit, where it can be shown that $<\theta_i^2>=3/4$, since
for all the cases considered the bound mesons have isospin 1/2. We denote this as HQS option.
Since it is not
yet settled which the best way to proceed at the charm mass scale is, both
cases will be considered here in order to determine the related uncertainty.
The second one has to do with the
way in which $< \vec R \cdot \vec \Theta>$ is calculated (see Ref.\cite{Oh:2007cr}). In fact,
this term might induce mixings between states with different $K$ and/or $k_+$. This effect will be neglected
in what follows.
The resulting formula for the rotational corrections to the mass of a system composed by a
soliton and two bound mesons (one in a state with $k_+$ and the other with $k_-$) is then
\begin{eqnarray}
M_{rot}(I, J, k_+, k_-, K) &=& {1 \over 2 \Omega} \, \Bigl\{ I (I
+ 1) + c_+ c_- \left[ K (K + 1) - k_+ (k_+ + 1) - k_- (k_- + 1)
\right] + \delta
\nonumber \\
& &
[J(J + 1) - K (K + 1) - I (I + 1) ]
\Bigl[ {c_+ + c_- \over 2} +
{c_+ - c_- \over 2} \, {k_+ (k_+ + 1) - k_- (k_- + 1)
\over K (K + 1)} \Bigr] \Bigr\}
\end{eqnarray}
where
\begin{eqnarray}
\delta = \Big\{
\begin{array}{cc}
  c_+^2\ k_+ (k_+ + 1) + c_-^2\ k_- (k_- + 1) & \mbox{BSA}\\
  & \\
  3/2  & \mbox{HQS}\\
\end{array}
\end{eqnarray}
Using the
BSA option one recovers the mass formula given in
Refs.\cite{Rho:1992yy,Riska:1991qd}.
Using the parameters of Table I the rotational corrections can then be
calculated. Note that in the case of
SMHQS scheme most of the hyperfine splittings needed for the calculation of
the rotational corrections have not been given in the literature.
Thus, only the predictions as obtained in the NSM
scheme are reported in Table III. The quoted values correspond to
the average between the results obtained using each of the two options
for the calculation of the rotational corrections. The corresponding uncertainty
is considered to be half of the difference between these two values, which turns
out to be about 70 MeV.
\begin{table}[h]
\begin{center}
\begin{tabular}{ccl}
  \hline
\hspace*{.2cm} $J^\pi$ \hspace*{.2cm} && Mass [GeV] \\
\\ \hline
 $1/2^+$ && 4.11, 4.14, 4.51, 4.57, 4.59\\
 $3/2^+$ && 4.16, 4.52, 4.60, 4.60 \\
 $1/2^-$ && 4.30, 4.35, 4.44 \\
 $3/2^-$ && 4.40,  4.43, 4.48 \\
 $5/2^-$ && 4.50 \\
 $5/2^+$ && 4.64 \\
 \hline
\end{tabular}
\end{center}
\caption{Masses including rotational corrections for the NSM scheme. All quoted values have an uncertainty
of about 70 MeV due to the ambiguities in the formula for the rotational corrections. The lower
value of this uncertainty range corresponds to the BSA option for the calculation of the rotational
corrections while the upper to the HQS one.}
\end{table}
From this table we see that the prediction for the mass of the lowest $3/2^-$ is in the range $4.33 -4.47$ GeV.
In the case of the $5/2^+$ the predicted mass is in the range $4.57- 4.71$ GeV.

\vspace*{2cm}

\section{Conclusions}

The utility of the bound state interpretation described above is of course hinged on the existence of a $C=-1$ meson bound state.
The calculations made within the framework of the NSM approximation indicate that such a state does exist.
The SMHQS approach is less definite on this issue, but most of the calculations
point towards the existence of a loosely bound state in the $C=-1$ channel. It should be noted that even if it is unbound
but lies close to threshold, one can still argue that an attractive $\bar D D$ interaction
can make the whole soliton-$\bar D$-$D$ system bound.

Given the assumption of the existence of a loosely $C=-1$ bound
state, the existence of a pentaquark-type state with quantum
numbers $(I,J^\pi)=(1/2, 3/2^-)$ and mass in a region compatible
with the LHCb observation follows naturally. Note that in the
present picture at least one of the components of this state comes
from populating the $k_+^{\pi_+} = 1/2^-$ meson state. In this
sense it appears as akin to the $\Lambda(1405)$. Note also that
according to HQS such a meson bound state should be degenerate
with a $k_+^{\pi_+} = 3/2^-$ state (see Fig.1). This probably then
explains the existence of the other two $3/2^-$ state close by.
The situation concerning the $(I,J^\pi)=(1/2, 5/2^+)$ state is
less clear. The model does predict such state but the associated
mass lies about 4.6 GeV, which is somewhat too high as compared
with the observed value. It has in fact been suggested that the
observed peak at 4450 MeV might even be a  kinematical effect
\cite{Guo:2015umn}. In any case it is important to recall that,
given the approximations made in the calculation of the masses,
the quoted values have to be viewed only as first estimates.

The present model obviously predicts the existence of several
other states. In particular, it predicts two $1/2^+$ and one
$3/2^+$ states with masses in the range $\simeq 4.1-4.2$ GeV,
which arise by putting the $C=+1$ meson in the lowest $k_+^{\pi_+}
= 1/2^+$ bound state. Whether those states are too wide to be
discriminated by the experiment remains to be seen.

This bound state approach picture can be fairly straightforwardly extended
to the bottom sector, where it would imply the existence also of hidden bottom pentaquarks
in view of the large mass of the bottom mesons. This would probably require a full
calculation within the SMHQS scheme, which is anyhow required
to obtain more accurate predictions for the properties of the states discussed in the
present work.

\begin{acknowledgments}
This work was partially supported by CONICET (Argentina) under
grant PIP 00682 and by ANPCyT (Argentina) under grant
PICT-2011-0113.
\end{acknowledgments}

\end{document}